\documentclass[fleqn,10pt]{wlscirep}
\usepackage[T1]{fontenc}

\usepackage{anyfontsize}
\usepackage{bm}
\usepackage{amsfonts}
\usepackage[utf8]{inputenc}
\usepackage{graphicx}
\usepackage{wrapfig}
\usepackage{epstopdf}
\usepackage{subfig}
\usepackage{setspace}
\usepackage{color}
\usepackage{siunitx}
\usepackage{fancyhdr}
\usepackage{comment}
\usepackage{enumitem}
\usepackage{setspace}
\graphicspath{ {FIGS/} }
\usepackage{array}
\usepackage{amsmath}
\usepackage{amssymb}
\usepackage{amsthm}
\usepackage{physics}
\usepackage{mathtools}
\usepackage{hyperref}
\usepackage{tikz}
\usetikzlibrary{positioning, calc, fit, arrows.meta, shapes.geometric, decorations.markings}

\definecolor{mylaser}{RGB}{245, 146, 54}    
\definecolor{mymod}{RGB}{105, 170, 255}     
\definecolor{mydet}{RGB}{53, 196, 168}      
\definecolor{myfiber}{RGB}{240, 140, 50}    
\definecolor{mypulsedark}{RGB}{60, 65, 75}  
\definecolor{optline}{RGB}{240, 165, 85}  
\definecolor{pmblue}{RGB}{115, 170, 255}  

\title{Enhancing the security of coherent one-way quantum key distribution using CHSH correlations}

\author[1,2,*]{Mahdi Shaban}
\author[2]{Farnaz Farman}
\author[1,2]{Alireza Bahrampour}
\affil[1]{Department of Physics, Sharif University of Technology, Tehran, Iran}
\affil[2]{Centre for Quantum Engineering and Photonics Technology, Sharif University of Technology, Tehran, Iran}

\affil[*]{mahdishaban20@gmail.com}

\begin{abstract}
The coherent one-way (COW) protocol is a quantum key distribution scheme
that has attracted significant attention, leading to the development
and commercialization of practical implementations.
Despite this progress, the security of the COW protocol has remained
a fundamental challenge since its introduction.
Numerous studies have investigated its security, and several security proofs
have been proposed over the years.
More recently, a number of works have questioned the security of this protocol.
In particular, one of the latest studies introduced an attack that severely
limits the security of COW-QKD and reported a maximum secure distance of
less than $20\,\mathrm{km}$.

In this work, we introduce minimal alteration to the COW protocol that
can enhance its security.
Specifically, instead of monitoring the coherence between successive pulses,
we propose to monitor quantum correlations through the violation of
Bell inequalities.
This approach enables the detection of a broader class of potential attacks.
Our simulation results indicate that, by employing this method,
the maximum secure distance of the protocol can be extended to
approximately $259\,\mathrm{km}$.
\end{abstract}

\begin{document}

\flushbottom
\maketitle
%
%
\thispagestyle{empty}


\section*{Introduction}

Quantum key distribution (QKD) \cite{bb84, e91} enables two distant parties to establish a shared secret key with security guaranteed by the laws of quantum mechanics. Since the seminal Bennett--Brassard 1984 (BB84) protocol \cite{bb84}, numerous QKD schemes have been proposed and experimentally demonstrated \cite{schem1,schem2,schem3}. Despite this progress, the security of practical QKD systems is inevitably affected by device imperfections, which may open security loopholes. One fundamental challenge arises from source imperfections, most notably photon-number-splitting (PNS) attacks \cite{pns}. Various countermeasures have therefore been proposed, including decoy-state techniques \cite{decoy1,decoy2}, strong reference-pulse schemes \cite{strong}, nonorthogonal coding \cite{nonorth}, and distributed-phase-reference (DPR) protocols \cite{dpr1,dpr2,dpr3,dpr4}. Notably, recent developments in this family include the small-number-random differential phase shift (SNRDPS) protocol, which enhances noise tolerance by introducing minimal randomness \cite{dps}.

Among DPR protocols, the coherent-one-way (COW) QKD protocol \cite{cow} has attracted significant attention due to its simple one-way architecture and experimental feasibility \cite{imp1,imp2,imp3,imp4,imp5,imp6,imp7,imp8,imp9}. COW-QKD has been implemented in field trials and deployed in operational quantum communication networks \cite{network}, highlighting its practical relevance. Under restricted classes of collective attacks, the secret key rate of COW-QKD scales linearly with the channel transmittance $\eta$ \cite{linear}. However, it was later shown that unconditional security of COW-QKD implies a quadratic scaling of the secret key rate, of order $\mathcal{O}(\eta^2)$ \cite{quad}. While recent studies have successfully established tight finite-key security bounds for COW-QKD variants within the composable security framework \cite{imp11}, most experimental demonstrations of COW-QKD to date \cite{imp5,imp6,imp7,imp8,imp9}, including long-distance implementations \cite{imp4,imp10}, still rely on security analyses predicting linear scaling, revealing a gap between practical performance and rigorous security guarantees.

A central difficulty in the security of COW-QKD is the so-called zero-error attack~\cite{zero}, in which an eavesdropper can extract information without introducing bit errors or destroying the coherence between adjacent nonvacuum pulses. Consequently, coherence monitoring alone is insufficient to guarantee security, and any analysis predicting a linear scaling of the key rate with $\eta$ is fundamentally insecure. More recently, Yin et al. \cite{simple} proposed a refined security analysis for COW-QKD based on a precise estimation of the phase error rate derived from detailed monitoring statistics. Their approach preserves the experimental simplicity of the original COW protocol and leads to an analytical lower bound on the secret key rate of $0.005\,\eta^2$, enabling secure key distribution over distances up to 100 km. Although this represents a substantial improvement over earlier unconditional security analyses \cite{simple}, extending the secure distance and strengthening the underlying security certification of COW-QKD remain important open challenges.

A conceptually different approach to addressing device imperfections is provided by device-independent quantum key distribution (DI) QKD \cite{di1,di2}. In DI-QKD, security is certified through the violation of Bell inequalities, without requiring assumptions about the internal functioning of the devices. Such strong guarantees, however, come at a high experimental cost, as they require loophole-free Bell tests with very high detection efficiencies. To bridge the gap between DI-QKD and prepare-and-measure protocols, semi-device-independent (SDI) QKD has been proposed \cite{sdi1,sdi2,sdi3}, where devices are treated as black boxes except for a bound on the dimension of the Hilbert space. This relaxation allows security to be established in a unidirectional configuration while still exploiting Bell-type correlations as a security resource.


In this work, we build on this semi-device-independent perspective to introduce a modified security framework for the COW-QKD protocol. Our enhancement involves two key modifications to the monitoring scheme: the inclusion of an additional decoy state and the replacement of the standard coherence check with a direct monitoring of Bell-type correlations via a CHSH test. These changes are designed to be implemented with minimal alteration to the existing experimental setup. A key merit of this work is the adaptation of the CHSH inequality—traditionally used in entanglement-based schemes—into a prepare-and-measure COW framework. This approach provides a versatile monitoring tool that could potentially be extended to other coherent-state-based protocols. Our robust security analysis shows that this revised framework restores a linear-like scaling behavior, marking a significant breakthrough by extending the secure distance from the previously reported sub-20 km limits to 259 km. This advancement substantially improves the practical viability of COW-QKD for long-haul networks.


\section*{Protocol description}
\subsection{original setup}

In the coherent-one-way (COW) QKD protocol, for each transmission round, Alice prepares and sends a signal state from the set $\{|0_z\rangle, |1_z\rangle, |0_x\rangle\}$. These signals consist of two optical time-bin pulses, each being either in the vacuum state $|0\rangle$ or a coherent state $|\alpha\rangle$ with mean photon number $\mu = |\alpha|^2$. Specifically, the data signals encoding bit values 0 and 1 are defined as $|0_z\rangle = |0\rangle|\alpha\rangle$ and $|1_z\rangle = |\alpha\rangle|0\rangle$, respectively. These are generated with equal \textit{a priori} probabilities $P(0_z) = P(1_z) = (1 - f)/2$. Additionally, a decoy signal $|0_x\rangle = |\alpha\rangle|\alpha\rangle$ is prepared with probability $P(0_x) = f$. Note that the temporal sequence of pulses is typically represented from right (early) to left (late).

At the receiver side, Bob employs an asymmetric beam splitter with transmittance $t_B$ to split the incoming signals into a data line and a monitoring line. The data line, equipped with a single-photon detector $D_B$, is utilized for raw key generation. Bob assigns a bit value 0 (1) if $D_B$ clicks in the first (second) time slot of a round. In the event of a double click, a random bit value is assigned. Conversely, the monitoring line features a Mach-Zehnder interferometer (MZI) followed by two detectors, $D_{M1}$ and $D_{M2}$, corresponding to constructive and destructive interference, respectively. This setup is designed to monitor the coherence between adjacent pulses; specifically, the interference condition ensures that two consecutive coherent states $|\alpha\rangle$ result in a click only at $D_{M1}$ and never at $D_{M2}$.
Following the quantum transmission phase, Bob publicly announces the rounds in which a detection event occurred at $D_B$. Alice then reveals which of these rounds corresponded to data signals ($|0_z\rangle$ or $|1_z\rangle$). The bits associated with these instances form the sifted key.

The security of the protocol relies on three key parameters: the gain $Q_z$, representing the probability of a detection in the data line; the quantum bit error rate (QBER); and the visibility, which quantifies the coherence between adjacent pulses. The visibility for a specific sequence $s$ is defined as:
\begin{equation}
	V_s = \frac{P(D_{M1}|s) - P(D_{M2}|s)}{P(D_{M1}|s) + P(D_{M2}|s)},
	\label{eq:visibility}
\end{equation}
where $s$ denotes a sequence of signals containing two adjacent coherent pulses. In terms of our defined states, the relevant sequences are $s \in \{0_x, 0_z1_z, 0_z0_x, 0_x1_z, 0_x0_x\}$. Here, $P(D_{Mi}|s)$ is the conditional probability of a click in detector $D_{Mi}$ given the sequence $s$. For instance, $V_{0_x}$ corresponds to the visibility within a single decoy signal $|0_x\rangle$, while $V_{0_z0_x}$ measures the coherence between the second pulse of a $|0_z\rangle$ signal and the first pulse of a subsequent $|0_x\rangle$ signal.

\subsection{Modifications}
To enhance the security analysis and enable the verification of Bell-type inequalities, we introduce specific modifications to both Alice's state preparation and Bob's measurement setup.

\textbf{Alice's Side:}
We extend the set of decoy states to include a fourth state, $|1_x\rangle$, making the protocol symmetric in the X-basis. As illustrated in the state definitions below, the X-basis states are formed by coherent superpositions that approximate the qubit superposition states in the subspace spanned by $|0_z\rangle$ and $|1_z\rangle$:
\begin{align}
	|0_z\rangle &= |0\rangle|\alpha\rangle, \nonumber \\
	|1_z\rangle &= |\alpha\rangle|0\rangle, \nonumber \\
	|0_x\rangle &= |\frac{\alpha}{\sqrt{2}}\rangle|\frac{\alpha}{\sqrt{2}}\rangle \sim \frac{|0_z\rangle + |1_z\rangle}{\sqrt{2}}, \\
	|1_x\rangle &= |-\frac{\alpha}{\sqrt{2}}\rangle|\frac{\alpha}{\sqrt{2}}\rangle \sim \frac{|0_z\rangle - |1_z\rangle}{\sqrt{2}}. \nonumber
\end{align}
where the approximation is valid for sufficiently small $\alpha$ (i.e., $\alpha \ll 1$). In this modified scheme, Alice chooses the transmission basis randomly: with probability $1-f$, she selects the Z-basis (sending $|0_z\rangle$ or $|1_z\rangle$), and with probability $f$, she selects the X-basis (sending $|0_x\rangle$ or $|1_x\rangle$). Within the chosen basis, the specific bit value (0 or 1) is selected with equal probability ($p=0.5$). Consequently, the probabilities for the decoy states are now symmetric: $P(0_x) = P(1_x) = f/2$.

\textbf{Bob's Side:}
The detection setup is adapted to check the CHSH inequality rather than simple visibility. The Data Line remains a Z-basis measurement ($\sigma_z$), distinguishing between early and late time-bins. The Monitoring Line, which traditionally serves as an X-basis measurement ($\sigma_x$) using a Mach-Zehnder interferometer (MZI), as described in~\cite{timebin}, is modified to allow for variable measurement bases.
Specifically, by adjusting the splitting ratio of the second beam splitter in the MZI to 85:15 and introducing a phase difference (0 or $\phi$) between the interferometer arms, Bob can rotate his measurement basis. This modification is essential for testing the CHSH inequality, defined by the parameter $S$:
\begin{equation}
	S = \langle A_1 B_1 \rangle + \langle A_1 B_2 \rangle + \langle A_2 B_1 \rangle - \langle A_2 B_2 \rangle.
\end{equation}
Here, Alice's observables are fixed as $A_1 = \sigma_z$ and $A_2 = \sigma_x$. To achieve the maximum violation of the CHSH inequality (Tsirelson's bound), Bob's observables must be set to:
\begin{equation}
	B_1 = \frac{\sigma_z + \sigma_x}{\sqrt{2}}, \quad B_2 = \frac{\sigma_z - \sigma_x}{\sqrt{2}}.
\end{equation}
The implementation details for realizing these specific bases ($B_1$ and $B_2$) within the monitoring line are discussed in the following section.

We now focus on the physical realization of the measurement bases $B_1 = \frac{1}{\sqrt{2}} (\sigma_z + \sigma_x)$ and $B_2 = \frac{1}{\sqrt{2}} (\sigma_z - \sigma_x)$ within the monitoring line. Since the security verification via the CHSH inequality is central to our modified protocol, these specific bases must be implemented in the monitoring branch.
While a standard asymmetric Mach-Zehnder interferometer (AMZI) with 50:50 beam splitters typically functions as a fixed X-basis measurement device, we can generalize its operation by exploiting its available degrees of freedom: relative intensity and relative phase. In the experimental setup, these correspond to the beam splitter splitting ratios and the phase shift induced by the modulator, respectively.
By modeling the entire interferometer as a unitary transformation, we can tune its components to shape the desired measurement bases. We approach this using a matrix representation of the optical components. We consider a general setup comprising a first beam splitter ($BS_1$) with transmittance and reflectance coefficients $(T_1, R_1)$, a phase modulator introducing a phase $\phi$ combined with a time delay, and a second beam splitter ($BS_2$) with coefficients $(T_2, R_2)$.

A dimensionality mismatch arises because the interferometer has two spatial modes at the input and output, but operates in the time domain. Specifically, an input consisting of two time-bins results in three time-bins at the output due to the delay line. This prevents a direct unitary description of the system using standard $2 \times 2$ matrices. To resolve this and derive a valid unitary operator for the entire interferometer, we introduce a virtual time mode at the input, expanding the Hilbert space. The detailed derivation of this mathematical treatment is provided in Appendix \ref{app:MZI_calc}.

Let $U_{MZ}$ denote the total transformation of the Mach-Zehnder interferometer. By composing the operations of the beam splitters and the delay/phase element ($U_{MZ} = BS_2 \cdot D \cdot BS_1$), we derive the following unitary matrix acting on the expanded space:

	\begin{equation}
		U_{MZ} =
		\begin{pmatrix}
			\sqrt{T_2 T_1} & i \sqrt{T_2 R_1} & 0 & 0 & -\sqrt{R_2 R_1}e^{i\phi} & i\sqrt{R_2 T_1}e^{i\phi} \\
			i\sqrt{R_2 T_1} & -\sqrt{R_2 R_1} & 0 & 0 & i\sqrt{T_2 R_1}e^{i\phi} & \sqrt{T_2 T_1}e^{i\phi} \\
			-\sqrt{R_2 R_1}e^{i\phi} & i\sqrt{R_2 T_1}e^{i\phi} & \sqrt{T_2 T_1} & i \sqrt{T_2 R_1} & 0 & 0 \\
			i\sqrt{T_2 R_1}e^{i\phi} & \sqrt{T_2 T_1}e^{i\phi} & i\sqrt{R_2 T_1} & -\sqrt{R_2 R_1} & 0 & 0 \\
			0 & 0 & -\sqrt{R_2 R_1}e^{i\phi} & i\sqrt{R_2 T_1}e^{i\phi} & \sqrt{T_2 T_1} & i \sqrt{T_2 R_1} \\
			0 & 0 & i\sqrt{T_2 R_1}e^{i\phi} & \sqrt{T_2 T_1}e^{i\phi} & i\sqrt{R_2 T_1} & -\sqrt{R_2 R_1}
		\end{pmatrix}
	\end{equation}

As shown, this transfer matrix is governed by three adjustable parameters: $T_1$, $T_2$, and $\phi$.

To physically realize the bases $B_1$ and $B_2$, we must ensure that the eigenstates of these operators, when injected into the interferometer, result in deterministic detection events (i.e., constructive interference at one detector and destructive at the other). It is crucial to note that the relevant interference occurs solely within the second time slot, where the "late" component of the short path overlaps with the "early" component of the long path.

By solving the interference conditions for the target eigenstates, we determine the required configuration for the interferometer components. Specifically, for the $B_1$ basis, the parameters must be set to $(T_1, R_1) = (0.5, 0.5)$, $(T_2, R_2) = (0.85, 0.15)$, and a phase shift of $\phi = 0$. For the $B_2$ basis, the splitting ratios remain the same with $(T_1, R_1) = (0.5, 0.5)$ and $(T_2, R_2) = (0.85, 0.15)$, while the phase shift is toggled to $\phi = \pi$.

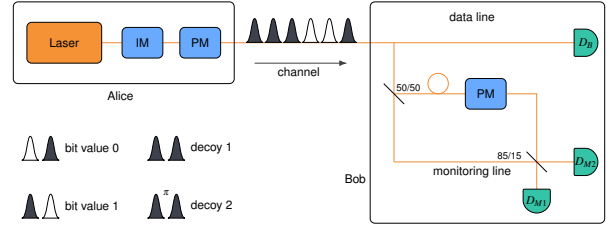
\begin{figure}[t]
	\centering
	
	\tikzset{
	}
	
	\resizebox{0.45\textwidth}{!}{
		\begin{tikzpicture}[
			font=\sffamily,
			box/.style={draw, thick, rounded corners=4pt},
			laser/.style={box, fill=mylaser, minimum width=2.2cm, minimum height=1.2cm},
			mod/.style={box, fill=mymod, minimum width=1.2cm, minimum height=0.9cm},
			fiber/.style={draw, thick, myfiber},		
			p_dark/.style={fill=mypulsedark},
			p_light/.style={fill=white},		
			gausspulse/.pic={
				\draw[thick, pic actions] 
				(-0.25, -0.35) 
				.. controls (-0.1, -0.35) and (-0.1, 0.35) .. (0, 0.35)
				.. controls (0.1, 0.35) and (0.1, -0.35) .. (0.25, -0.35) 
				-- cycle;}
			]
			
			\draw[fiber] (-4.5, 0) -- (10.5, 0);
			
			\node[laser] (laser) at (-4.5, 0) {Laser};
			\node[mod] (im) at (-2.2, 0) {IM};
			\node[mod] (pm1) at (-0.5, 0) {PM};
			
			\draw[thick, rounded corners=4pt] (-6.0, 1.2) rectangle (0.5, -1.2);
			\node[below] at (-2.85, -1.3) {Alice};
			
			\draw[-{Latex[length=2mm]}, draw=darkgray, thin] (1.1, -0.6) -- (3.8, -0.6);
			\node[below] at (2.4, -0.6) {channel};
			
			\pic[p_dark]  at (1.1, 0.36) {gausspulse};
			\pic[p_dark]  at (1.65, 0.36) {gausspulse};
			\pic[p_dark]  at (2.2, 0.36) {gausspulse};
			\pic[p_light] at (2.75, 0.36) {gausspulse};
			\pic[p_light] at (3.3, 0.36) {gausspulse};
			\pic[p_dark]  at (3.85, 0.36) {gausspulse};
			
			\draw[thick, rounded corners=4pt] (4.5, 1.2) rectangle (11.4, -5.3);
			\node[left] at (4.5, -4.0) {Bob};
			
			\draw[fiber] (5.2, 0) -- (5.2, -3.5);
			\draw[fiber] (5.2, -1.5) -- (8.5, -1.5);
			\draw[fiber] (5.2, -3.5) -- (10.5, -3.5);
			
			\draw[fiber] (6.5, -1.2) circle (0.3);
			
			\node[mod] (pm2) at (7.9, -1.5) {PM};
			\draw[fiber] (pm2.east) -- (9.4, -1.5) -- (9.4, -4.3);
			
			\draw[thick] (4.9, -1.2) -- (5.5, -1.8);
			\node[above right, inner sep=2pt] at (5.2, -1.5) {\footnotesize 50/50};
			
			\draw[thick] (9.1, -3.2) -- (9.7, -3.8);
			\node at (8.6, -3.3) {\footnotesize 85/15};
			
			\draw[thick, fill=mydet, rounded corners=1pt] (10.5,-0.4) -- (10.8,-0.4) arc(-90:90:0.4) -- (10.5,0.4) -- cycle;
			\node at (10.85, 0) {$D_B$};
			
			\draw[thick, fill=mydet, rounded corners=1pt] (10.5,-3.9) -- (10.8,-3.9) arc(-90:90:0.4) -- (10.5,-3.1) -- cycle;
			\node at (10.85, -3.5) {$D_{M2}$};
			
			\draw[thick, fill=mydet, rounded corners=1pt] (9.0,-4.3) -- (9.0,-4.6) arc(180:360:0.4) -- (9.8,-4.3) -- cycle;
			\node at (9.4, -4.65) {$D_{M1}$};
			
			\node[above] at (7.5, 0.5) {data line};
			\node[below] at (7.5, -3.5) {monitoring line};
			
			\newcommand{\pulsepair}[4]{
				\pic[#3] at (#1, #2) {gausspulse};
				\pic[#4] at (#1+0.6, #2) {gausspulse};
			}
			
			\pulsepair{-5.5}{-3.1}{p_light}{p_dark}
			\node[right] at (-4.6, -3.1) {bit value 0};
			
			\pulsepair{-5.5}{-4.8}{p_dark}{p_light}
			\node[right] at (-4.6, -4.8) {bit value 1};
			
			\pulsepair{-1.8}{-3.1}{p_dark}{p_dark}
			\node[right] at (-0.9, -3.1) {decoy 1};
			
			\pulsepair{-1.8}{-4.8}{p_dark}{p_dark}
			\node[above] at (-1.5, -4.6) {\footnotesize $\pi$};
			\node[right] at (-0.9, -4.8) {decoy 2};
			
		\end{tikzpicture}
	}
	
	\caption{Schematic of the modified COW protocol considered in our work. Alice randomly sends a sequence of states $|0\rangle_{2k-1} |\alpha\rangle_{2k}$, $|\alpha\rangle_{2k-1} |0\rangle_{2k}$, $|\frac{\alpha}{\sqrt{2}}\rangle_{2k-1} |\frac{\alpha}{\sqrt{2}}\rangle_{2k}$, and $|-\frac{\alpha}{\sqrt{2}}\rangle_{2k-1} |\frac{\alpha}{\sqrt{2}}\rangle_{2k}$ to Bob. Bob routes all incoming pulses to the data line with probability $t_B$, and to the monitoring line with probability $1 - t_B$. Compared to the original version of the protocol, Alice sends an additional decoy state. Moreover, in Bob’s Mach–Zehnder interferometer, the splitting ratio of one beam splitter is modified to 85/15, and a phase modulator is inserted in one of its arms. $D_B$, $D_{M_1}$, and $D_{M_2}$ are the single-photon detectors.}
	\label{fig:cow_setup}
\end{figure}

\subsection{Modified Protocol}
Building upon the modifications detailed in the previous section, we now present the proposed protocol. This design aims to enhance the security of the scheme and improve its capability to detect attacks compared to prior methods.

Alice encodes her logical bits into sequences of optical pulses. The signal states, used for key extraction, are defined as:
\begin{equation}
	\ket{0} = \ket{0}\ket{\alpha}, \quad \ket{1} = \ket{\alpha}\ket{0}
\end{equation}
where $\ket{\alpha}$ represents a coherent state with a mean photon number of $\mu = |\alpha|^2$. Each signal state is transmitted with a probability of $(1-f)/2$. In addition to the signal states, Alice transmits decoy states defined as:
\begin{equation}
	\ket{f_1} = \ket{\frac{\alpha}{\sqrt{2}}}\ket{\frac{\alpha}{\sqrt{2}}}, \quad \ket{f_2} = \ket{\frac{-\alpha}{\sqrt{2}}}\ket{\frac{\alpha}{\sqrt{2}}}
\end{equation}
Each decoy state is transmitted with a probability of $f/2$.

The schematic of the modified protocol is illustrated in Fig.~\ref{fig:cow_setup}. Alice prepares the required states using an Intensity Modulator (IM) to adjust the pulse intensity (selecting from $\{0, \frac{\alpha}{\sqrt{2}}, \alpha\}$) and a Phase Modulator (PM) to adjust the phase (selecting between $0$ and $\pi$). The pulses are then transmitted through a fiber channel with transmittance $t = 10^{-0.02L}$ to Bob.

Upon receiving the pulses, Bob splits the incoming beam into two paths using an asymmetric beam splitter with transmittance $t_B$: the \textit{Data Line} and the \textit{Monitoring Line}. In the Data Line, Bob employs a single-photon detector to distinguish whether a photon arrives in the first or second time window, thereby measuring the logical bits $0$ and $1$. The Monitoring Line consists of a non-symmetric Mach-Zehnder interferometer (AMZI) and two single-photon detectors. As specified in our design, the interferometer comprises a 50:50 beam splitter ($BS_1$) and an 85:15 beam splitter ($BS_2$). One arm contains a time delay $\Delta t$ (corresponding to the time interval between two pulses) and a PM for basis selection. By setting the PM phase to $0$ or $\pi$, Bob measures in the basis $B_1 = (\sigma_z + \sigma_x)/\sqrt{2}$ or $B_2 = (\sigma_z - \sigma_x)/\sqrt{2}$, respectively, and records the detector click statistics.

Following the transmission, Alice and Bob perform the sifting process. Alice publicly announces the positions of the decoy states. Bob discards the decoy states in the Data Line but utilizes them in the Monitoring Line. Furthermore, Alice reveals a fraction of the signal bits for error estimation. Using the revealed data, Bob calculates the CHSH inequality parameter $S$ in the Monitoring Line and the Quantum Bit Error Rate (QBER), denoted as $Q$, in the Data Line. The protocol is considered secure and proceeds to key extraction only if the conditions $S \ge S_0$ and $Q \le Q_0$ are met, where $S_0$ and $Q_0$ are pre-determined security thresholds tailored to the implementation conditions. Finally, classical post-processing is performed, including error correction to reconcile Alice and Bob's shared bits, and privacy amplification to minimize any potential information gained by an eavesdropper.

\section*{Security Analysis}
\subsection{secret key rate formula}
In this section, we provide a detailed security analysis of the proposed protocol. We begin by defining the extractable secret key length in the finite-size regime using the Quantum Leftover Hashing Lemma~\cite{qlhl} and then derive the asymptotic key rate.

Let $\boldsymbol{Z}_A$ denote the raw key string generated by Alice, and $\boldsymbol{E}'$ represent all the information possessed by the eavesdropper (Eve) up to the error-correction and error-verification steps. The uncertainty of Eve about the raw key is characterized by the smooth min-entropy, $H_{\min}^{\bar{\epsilon}}(\boldsymbol{Z}_A|\boldsymbol{E}')$, which quantifies the probability that Eve can guess $\boldsymbol{Z}_A$ using the optimal strategy~\cite{operational}. According to the quantum leftover hashing lemma, a $\epsilon_{\mathrm{sec}}$-secret key of length $l$ can be extracted from $\boldsymbol{Z}_A$ by applying a random universal$_2$ hash function, provided that the parameter $\epsilon_{\mathrm{sec}}$ satisfies:
\begin{equation}
	\epsilon_{\mathrm{sec}} = 2\bar{\epsilon} + \frac{1}{2}\sqrt{2^{l - H_{\min}^{\bar{\epsilon}}(\boldsymbol{Z}_A|\boldsymbol{E}')}}
\end{equation}
where $\bar{\epsilon}$ is the smoothing parameter. By setting $\epsilon_{\mathrm{PA}} = \frac{1}{2}\sqrt{2^{l - H_{\min}^{\bar{\epsilon}}(\boldsymbol{Z}_A|\boldsymbol{E}')}}$, the length $l$ of the secret key is given by:
\begin{equation}
	l = H_{\min}^{\bar{\epsilon}}(\boldsymbol{Z}_A|\boldsymbol{E}') - 2\log_2 \left( \frac{1}{2\epsilon_{\mathrm{PA}}} \right)
\end{equation}

for $\bar{\epsilon} + \epsilon_{\mathrm{PA}} \leq \epsilon_{\mathrm{sec}}$ (proof is by the quantum leftover hash lemma). During the post-processing stage, information is leaked to Eve for error correction and verification. Let $\text{leak}_{\text{EC}}$ be the amount of information revealed. A chain-rule inequality for smooth entropies allows us to bound the entropy after error correction:
\begin{equation}
	H_{\min}^{\bar{\epsilon}}(\boldsymbol{Z}_A|\boldsymbol{E}') \ge H_{\min}^{\bar{\epsilon}}(\boldsymbol{Z}_A|\boldsymbol{E}) - \text{leak}_{\text{EC}} - \log_2 \frac{2}{\epsilon_{\mathrm{cor}}}
	\label{eq:chain-rule}
\end{equation}
The QKD protocol is said to be $\varepsilon_{\mathrm{qkd}}$-secure, with
$\varepsilon_{\mathrm{qkd}} \ge \varepsilon_{\mathrm{cor}} + \varepsilon_{\mathrm{sec}}$,
if it is correct with probability at least $1 - \varepsilon_{\mathrm{cor}}$
and secret with probability at least $1 - \varepsilon_{\mathrm{sec}}$. Under the assumption of independent and identically distributed (i.i.d.) collective attacks, the joint state of Alice, Bob, and Eve, $\rho_{ABE}^{\otimes N}$, has a tensor product structure. In this case, the smooth min-entropy of the total state is bounded by the min-entropy of a single transmission:
\begin{equation}
	H_{\min}^{\bar{\epsilon}}(\boldsymbol{Z}_A|\boldsymbol{E}) \ge H_{\min}^{\epsilon}(\rho_{Z_A E}^{\otimes n} | \rho_E^{\otimes n})
\end{equation}
where $\bar{\epsilon} \ge \epsilon $. The term $\text{leak}_{\text{EC}}$ is upper-bounded by the inefficiency of the error correction code ($f_{\text{EC}}$) and the entropy of $Z_A$ conditioned on $Z_B$. While in the finite regime this includes logarithmic correction terms, it is primarily driven by $n f_{\text{EC}} H(Z_A|Z_B)$.

To compute the final key rate, we consider the asymptotic limit where the number of signals $n \to \infty$. In this regime, we use the property that the smooth min-entropy converges to the standard min-entropy as the smoothing parameter $\epsilon \to 0$:
\begin{equation}
	H_{\min}^{\epsilon}(\rho_{Z_A E}^{\otimes n} | \rho_E^{\otimes n}) \xrightarrow{\epsilon \to 0} H_{\min}(\rho_{Z_A E}^{\otimes n} | \rho_E^{\otimes n})
\end{equation}
Then, using the additivity of min-entropy (derived in Lemma 3.1.6 of ~\cite{renner}) we have that:
\begin{align}
	H_{\min}(\rho_{Z_A E}^{\otimes n} | \rho_E^{\otimes n}) &= n H_{\min}(\rho_{Z_A E} | \rho_E) \nonumber \\
	&= n H_{\min}(Z_A|E)
	\label{eq:hmin}
\end{align}
Consequently, the asymptotic secret key rate per transmission, $R = \lim_{N \to \infty} \frac{l}{N}$, is derived by substituting Eqs.~\eqref{eq:chain-rule} and~\eqref{eq:hmin} into the key length formula and neglecting the terms that vanish as $N \to \infty$ (such as $\frac{1}{N}\log_2(1/\epsilon)$ terms). The rate is thus given by:
\begin{equation}
	R \ge \frac{n}{N} \, \lbrack H_{\min}(A|E) - \frac{1}{n}\text{leak}_{\text{EC}} \rbrack
	\label{eq:rate}
\end{equation}
where $\frac{1}{n} \text{leak}_{\text{EC}} \le f_{\text{EC}} h(E_z)$, and $h(E_z)$ is the binary entropy function and $E_z$ is the bit error rate in the Z basis. The pre-factor $\frac{n}{N}$ corresponds to the Z-basis gain, denoted as $Q_z$, which represents the probability that a signal prepared by Alice in the Z-basis is successfully detected by Bob on the data line.

Finally, to evaluate $H_{\min}(A|E)$ for our specific protocol, we utilize the CHSH correlation parameter $S$. It has been proven in Ref.~\cite{sdi3} that the min-entropy is lower-bounded by a function of the Bell parameter $S$ as follows:
\begin{equation}
	H_{\min}(A|E) \ge 1 - \log_2  (1 + \sqrt{2 - S^2/4} )
\end{equation}
This relation connects the secrecy of the key directly to the violation of the CHSH inequality. By substituting this bound into Eqs.~\eqref{eq:rate}, we obtain the final security formula for our modified protocol:
\begin{equation}
	R \ge Q_z \, \lbrack 1 - \log_2 ( 1 + \sqrt{2 - S^2/4} ) - f_{\text{EC}} h(E_z) \rbrack
\end{equation}
This equation confirms that for $S > 2$ (violation of local realism) and sufficiently low QBER ($E_z$), a positive secret key rate is achievable.

\subsection{CHSH parameter S}

\begin{figure}[t]
	\centering
	\includegraphics[width=0.5\textwidth]{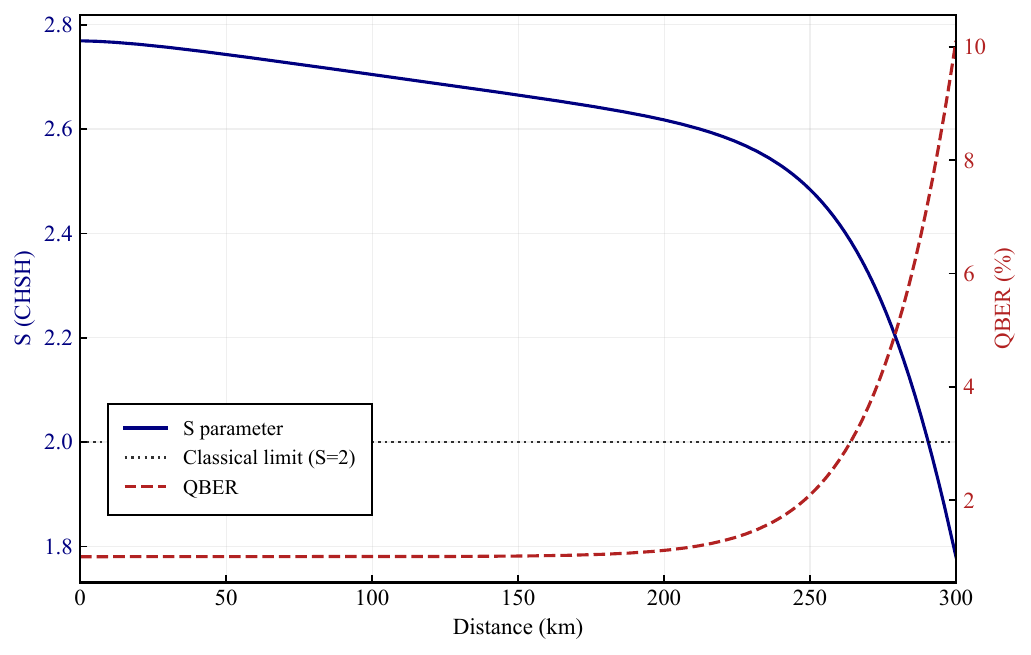}
	\caption{The CHSH inequality parameter $S$ (solid line) and the quantum bit error rate (dashed line) as a function of the transmission distance between Alice and Bob. The optical misalignment error is set to  $e_a =1\%$. The classical limit $S=2$ is indicated by the black dotted line.}
	\label{fig:QS}
\end{figure}
To test for non-local correlations and bound Eve's information, we employ the Clauser-Horne-Shimony-Holt (CHSH) inequality. Alice and Bob use their randomly chosen basis settings and corresponding measurement outcomes to compute the CHSH parameter, $S$. The terms $\langle A_i B_j \rangle$ within the CHSH expression represent the expectation values (correlations) of the outcomes for the measurement settings $A_i$ and $B_j$, where $i, j \in \{1, 2\}$.

Experimentally, these expectation values are determined from the detector click statistics. The correlation for a given pair of bases $(i, j)$ is calculated as:
\begin{equation}
	\langle A_i B_j \rangle = \frac{N_{ij}^{++} - N_{ij}^{+-} - N_{ij}^{-+} + 	N_{ij}^{--}}{N_{ij}^{++} + N_{ij}^{+-} + N_{ij}^{-+} + N_{ij}^{--}}.
	\label{eq:ab_experimental}
\end{equation}

Here, $N_{ij}^{ab}$ denotes the number of detection events where Alice sends a state corresponding to outcome $a \in \{+1, -1\}$ in basis $A_i$, and Bob measures an outcome $b \in \{+1, -1\}$ in basis $B_j$. In our setup, a click at detector $DM_1$ corresponds to the $+1$ eigenvalue, while a click at $DM_2$ corresponds to the $-1$ eigenvalue.	
We now turn to the ideal theoretical value of the $S$ parameter for our protocol. The detailed calculations for each correlation term, $\langle A_i B_j \rangle$, are derived from the quantum-mechanical detection probabilities within the asymmetric Mach-Zehnder interferometer described in the previous section. As shown in Appendix~\ref{app:CHSH_calc}, substituting the resulting theoretical values into the CHSH expression yields $S = 2\sqrt{2}$. This value represents the maximal violation of the CHSH inequality allowed by quantum mechanics (the Tsirelson bound), confirming the presence of perfect quantum correlations in the ideal case.

\section*{simulation Results}
To evaluate the performance of the modified protocol and compare it with the original version, a practical implementation or a rigorous simulation is required. Given the available resources, a simulation was conducted using Python to demonstrate the general performance of the modified protocol. In this simulation, the quantum channel (optical fiber) is modeled as a beam splitter with a transmittance equal to that of the physical channel. The channel misalignment is simplified and assumed to be constant along the channel length. This error is set to $1\%$ in the Data Line, meaning that the channel flips a bit $0$ to $1$ (and vice versa) with a probability of $0.01$. In the Monitoring Line, this misalignment results in a reduction of coherence between the pulses.
\begin{figure}[t]
	\centering
	\includegraphics[width=0.445\textwidth]{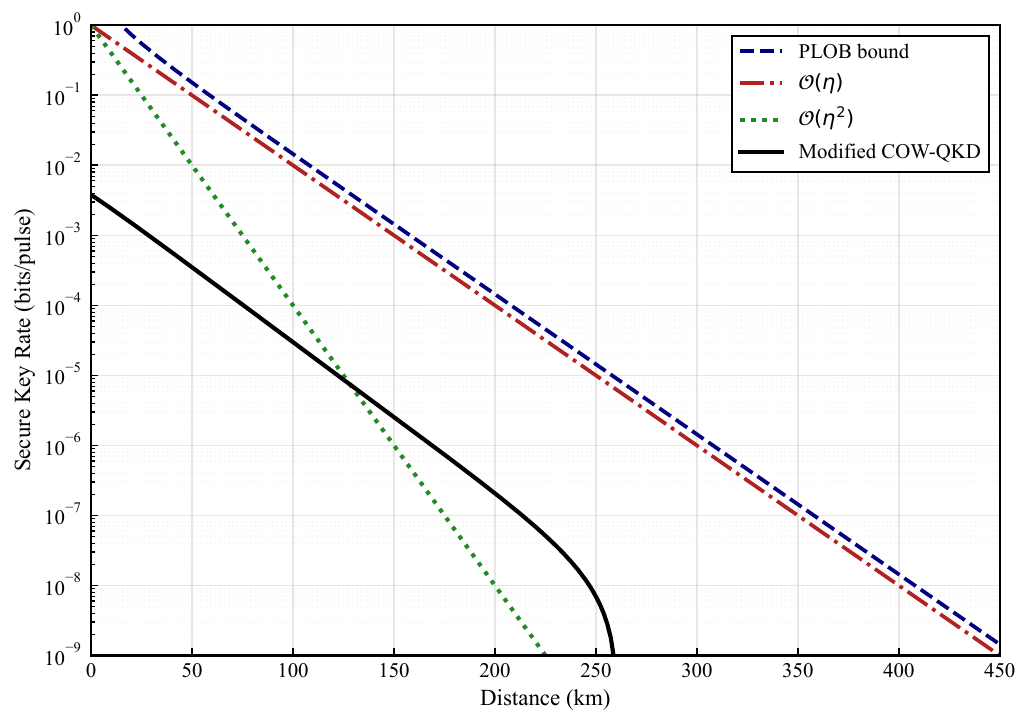}
	\caption{Secret key rates in the asymptotic regime. When $e_a = 1\%$, the secret key rate exhibits linear scaling, $\mathcal{O}(\eta)$. For comparison, the figure also shows reference curves with linear and quadratic scaling in $\eta$, together with the PLOB bound.}
	\label{fig:skr}
\end{figure}

We assume identical single-photon detectors for both parties. The detection of photons from coherent states is modeled using Poissonian statistics. The detectors are characterized by a quantum efficiency of $90\%$ and a dark count probability of $10^{-8}$. The error correction efficiency is set to $f_{EC} = 1.1$. The behavior of the Quantum Bit Error Rate (QBER), denoted as $Q$, and the CHSH parameter, $S$, as a function of channel length is illustrated in Fig.~\ref{fig:QS}.

The primary limiting factor in this protocol is the detector dark count rate. An analysis of the error rate and the parameter $S$ reveals that at long distances, the probability of a photon reaching the detector becomes negligible. Consequently, dark counts begin to dominate, causing a sharp increase in the error contribution within both the Data and Monitoring lines.

Using the derived values of $Q$ and $S$, the secure key rate can be calculated. As illustrated in Fig.~\ref{fig:skr}, the secure key rate decreases with a steeper slope beyond a certain distance, which is directly attributed to the impact of dark counts. From the slope of the curve, it can be deduced that the secure key rate scales linearly with the channel transmittance ($R \approx 0.002 \eta$). This linear scaling demonstrates a significant improvement in security compared to the original version of the protocol. Furthermore, the simulation results indicate that the maximum secure distance of the proposed protocol reaches $259$ km.

\begin{figure}[t]
	\centering
	\includegraphics[width=0.5\textwidth]{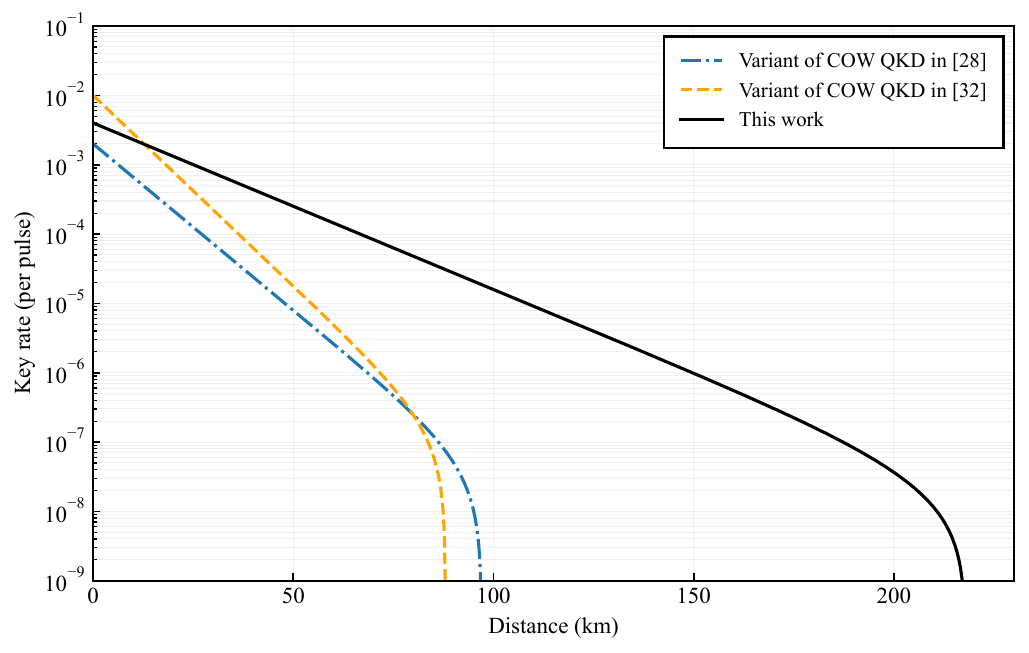}
	\caption{Comparison of asymptotic secret key rates of COW-QKD in this work and the variants of COW-QKD using active basis choice~\cite{quad,simple}. The misalignment error $e_a$ is set to $1\%$, the dark count rate is set to $10^{-7}$, and the detection efficiency of $\eta_d = 99\%$. The variant in Ref.~\cite{quad} uses 6 optical pulses in each signal block and all blocks of signals share a common phase. The variant in Ref.~\cite{simple} adds a two-pulse vacuum state  $\lvert 0\rangle_{2k-1}\lvert 0\rangle_{2k}$ as decoy sequence 2.}
	\label{fig:comparison}
\end{figure}

To benchmark the results of this work against other published works on COW protocol security, the simulation must be conducted under identical conditions. Therefore, for the comparison phase, we adjust the detector specifications to a quantum efficiency of $99\%$ and a dark count probability of $10^{-7}$. Since the quantum channel modeling is consistent across these studies, a valid comparison is possible. The secure key rate as a function of distance is plotted in Fig.~\ref{fig:comparison}. This figure includes results from the work in Ref.~\cite{quad}, where coherent states without phase randomization are treated as blocks of three in-phase pulses. It also includes results from Ref.~\cite{simple}, where a vacuum decoy state $|0\rangle$ was added to the standard decoy state $|\alpha\rangle$. As evident from the graph, in the two previous studies, the secure key rate scales with the square of the channel transmittance ($\eta^2$), implying lower security and performance at long distances. In contrast, in the current work, the secure rate scales linearly with transmittance ($\eta$), offering superior performance at extended ranges. It is worth noting that the secure key rate of the Decoy-State BB84 protocol also scales linearly with channel transmittance.


\section*{Conclusion}
In this study, we propose slight modifications to the existing protocol to significantly improve its security. Specifically, we incorporate an additional decoy state on Alice’s side and modify the Mach-Zehnder interferometer within Bob’s monitoring line. Ultimately, we derive an asymptotic security proof for COW-QKD. Unlike previous approaches, our proof no longer relies on the coherence between adjacent pulses for eavesdropper detection; rather, it is based on the inter-pulse correlations quantified by the CHSH inequality. notably, our findings reveal that the secure key rate scales linearly with channel transmittance $\mathcal{O}(\eta)$, overcoming the quadratic dependence reported in prior security proofs. The significance of this work lies not only in the record-breaking extension of the secure distance to 259 km but also in introducing a robust methodology for monitoring prepare-and-measure protocols via Bell-type correlations. This framework opens new avenues for enhancing the security of various other QKD schemes with minimal hardware changes. By maintaining the general framework of the setup, we successfully pave the way for the practical and secure implementation of long-distance COW-QKD.


\appendix

\section{Derivation of Measurement Bases in the Asymmetric MZI}
\label{app:MZI_calc}

We aim to realize the two measurement bases $B_1 = \frac{1}{\sqrt{2}} (\sigma_z + \sigma_x)$ and $B_2 = \frac{1}{\sqrt{2}} (\sigma_z - \sigma_x)$ using an asymmetric Mach-Zehnder Interferometer (MZI). Since a Bell inequality test is required for monitoring the protocol, these two bases must be implemented in the monitoring line. As previously described, a standard MZI with two 50/50 beam splitters and a delay line acts as a measurement in the $X$ basis. We have two degrees of freedom at our disposal: the relative intensity and the relative phase between the arms. In the MZI, these correspond to the beam splitter's splitting ratio (relative intensity) and the phase modulator's applied phase (relative phase). By adjusting these parameters, any desired basis can be constructed for time-bin qubits.

If the entire interferometer is treated as a unitary transformation, we can configure its components to form our desired bases. We proceed by using a matrix representation for the MZI components. We consider a general setup with a first beam splitter with transmittance and reflectance $(T_1,R_1)$, a phase modulator applying a phase $\phi$ in one arm, which also contains the time delay, and a second beam splitter with parameters $(T_2,R_2)$.

A challenge arises from the mode transformation. The MZI has two spatial input and output modes. If we consider two input time modes (early and late), three time modes will be present at the output, which prevents the overall transformation from being unitary. To address this, we introduce a virtual third time mode at the input. This allows us to define a unitary transformation for the entire system.

\begin{figure}[t]
	\centering
	\resizebox{0.45\textwidth}{!}{
		\begin{tikzpicture}[
			beam_in/.style={
				thick, color=myfiber,
				decoration={markings, mark=at position 0.92 with {\arrow{Latex[length=2.5mm, width=2mm]}}},
				postaction={decorate}
			},
			beam_out/.style={
				thick, color=myfiber,
				decoration={markings, mark=at position 0.15 with {\arrow{Latex[length=2.5mm, width=2mm]}}},
				postaction={decorate}
			},
			beam_nofl/.style={ thick, color=myfiber }
			]
			
			
			\newcommand{\pulseup}[3]{
				\begin{scope}[shift={(#1,#2)}]
					\draw[thick, fill=white] 
					(-0.6, 0) 
					.. controls (-0.3, 0) and (-0.15, 0.9) .. (0, 0.9) 
					.. controls (0.15, 0.9) and (0.3, 0) .. (0.6, 0) 
					-- cycle;
					\node[font=\sffamily\normalsize, text=black] at (0, 0.35) {#3};
				\end{scope}
			}
			
			\newcommand{\pulseright}[3]{
				\begin{scope}[shift={(#1,#2)}]
					\draw[thick, fill=white] 
					(0, -0.6) 
					.. controls (0, -0.3) and (0.9, -0.15) .. (0.9, 0) 
					.. controls (0.9, 0.15) and (0, 0.3) .. (0, 0.6) 
					-- cycle;
					\node[font=\sffamily\normalsize, text=black] at (0.35, 0) {#3};
				\end{scope}
			}
			
			
			\draw[beam_in] (-4.5, 0) -- (0, 0);       
			\draw[beam_in] (0, 4.5) -- (0, 0);        
			
			\draw[beam_out] (8, -3.0) -- (12.5, -3.0);    
			\draw[beam_out] (8, -3.0) -- (8, -7.5);     
			
			\draw[beam_nofl] (0, 0) -- (8, 0) -- (8, -3.0);    
			\draw[beam_nofl] (0, 0) -- (0, -3.0) -- (8, -3.0);   
			
			\draw[thick, color=myfiber] (3.5, 0.6) circle (0.6);
			
			
			\draw[thick] (-0.7, 0.7) -- (0.7, -0.7);
			\node[font=\sffamily\small] at (0.7, 0.3) {$(T_1, R_1)$};
			
			\draw[thick] (7.3, -2.3) -- (8.7, -3.8);
			\node[font=\sffamily\small] at (6.7, -2.8) {$(T_2, R_2)$};
			
			\node[draw, thick, fill=mymod, rounded corners=3pt, 
			minimum width=1.8cm, minimum height=1.3cm, font=\sffamily] 
			at (6.10, 0) {PM};
			
			
			\pulseup{-1.5}{0}{$\mathsf{a'_1}$}
			\pulseup{-2.75}{0}{$\mathsf{a'_2}$}
			\pulseup{-4}{0}{$\mathsf{a'_3}$}
			
			\pulseright{0}{1.5}{$\mathsf{a_1}$}
			\pulseright{0}{2.75}{$\mathsf{a_2}$}
			\pulseright{0}{4}{$\mathsf{a_3}$}
			
			\pulseup{9.5}{-3.0}{$\mathsf{b_3}$}
			\pulseup{10.75}{-3.0}{$\mathsf{b_2}$}
			\pulseup{12}{-3.0}{$\mathsf{b_1}$}
			
			\pulseright{8}{-4.5}{$\mathsf{b'_3}$}
			\pulseright{8}{-5.75}{$\mathsf{b'_2}$}
			\pulseright{8}{-7.0}{$\mathsf{b'_1}$}
			
		\end{tikzpicture}
	}
	\caption{Input and output modes of the asymmetric Mach-Zehnder interferometer.}
	\label{fig:mzi}
\end{figure}
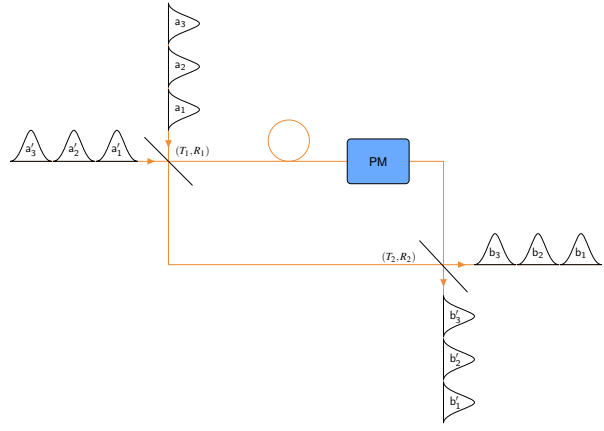

As shown in Fig.~\ref{fig:mzi}, we have a total of six input modes: time modes $a_1, a_2, a_3$ for spatial input 1, and time modes $a'_1, a'_2, a'_3$ for spatial input 2. Similarly, we have six output modes labeled $b_1, b_2, b_3$ for spatial output 1 and $b'_1, b'_2, b'_3$ for spatial output 2. In vector notation, the transformation is:
\begin{align}
	\begin{pmatrix}
		a_1 \\
		a'_1 \\
		a_2 \\
		a'_2 \\
		a_3 \\
		a'_3
	\end{pmatrix}
	\rightarrow
	\begin{pmatrix}
		b_1 \\
		b'_1 \\
		b_2 \\
		b'_2 \\
		b_3 \\
		b'_3
	\end{pmatrix}
\end{align}
Here, the first two elements of the vector correspond to the first time mode in the two different spatial modes, the next two elements correspond to the second time mode, and so on.
To ensure mathematical clarity regarding the operators and state vectors presented in the main text of the manuscript, we explicitly define the underlying vector space and the choice of basis.
The state space of the system is represented by a six-dimensional complex vector space (Hilbert space) $\mathcal{V}$, which is spanned by the set of orthonormal basis states $\{|i,j\rangle\}$:
\begin{equation}
	\mathcal{V} = \text{span}\left( \left\{ |i,j\rangle \right\} \right) = \left\{ \sum_{i=1}^{3} \sum_{j=1}^{2} \alpha_{i,j} |i,j\rangle \;\middle|\; \alpha_{i,j} \in \mathbb{C} \right\},
\end{equation}
where $i \in \{1,2,3\}$ represents the time mode (first, second, and third time bins, respectively), and $j \in \{1,2\}$ denotes the spatial mode (with $j=1$ corresponding to spatial input 1, and $j=2$ corresponding to spatial input 2).

We order the basis states in a time-major, spatial-minor sequence. The explicit representation of these basis states as six-dimensional column vectors is given by:
\begin{align}
	|1,1\rangle &= \begin{pmatrix} 1 & 0 & 0 & 0 & 0 & 0 \end{pmatrix}^T, \quad
	|1,2\rangle = \begin{pmatrix} 0 & 1 & 0 & 0 & 0 & 0 \end{pmatrix}^T, \nonumber \\
	|2,1\rangle &= \begin{pmatrix} 0 & 0 & 1 & 0 & 0 & 0 \end{pmatrix}^T, \quad
	|2,2\rangle = \begin{pmatrix} 0 & 0 & 0 & 1 & 0 & 0 \end{pmatrix}^T, \nonumber \\
	|3,1\rangle &= \begin{pmatrix} 0 & 0 & 0 & 0 & 1 & 0 \end{pmatrix}^T, \quad
	|3,2\rangle = \begin{pmatrix} 0 & 0 & 0 & 0 & 0 & 1 \end{pmatrix}^T.
\end{align}

Consequently, the operator for the first beam splitter with an arbitrary ratio $(T_1,R_1)$ is represented as:
\begin{align}
	BS_1 = 
	\begin{pmatrix}
		BS_{2\times 2} & 0_{2\times 2} & 0_{2\times 2} \\
		0_{2\times 2} & BS_{2\times 2} & 0_{2\times 2} \\
		0_{2\times 2} & 0_{2\times 2} & BS_{2\times 2} \\
	\end{pmatrix}
	,\\ \text{where} \quad BS_{2\times 2} =
	\begin{pmatrix}
		\sqrt{T_1} & i\sqrt{R_1} \\
		i\sqrt{R_1} & \sqrt{T_1}
	\end{pmatrix}
	\label{eq:BS} 
\end{align}
As expected, the $BS_1$ operator is block-diagonal, as it acts only on the spatial modes for each time mode independently and does not mix different time modes.

The operation of the time delay (by one time-bin interval) and the application of an arbitrary phase $\phi$ is given by the matrix $D$:
\begin{align}
	D =
	\begin{pmatrix}
		1 & 0 & 0 & 0 & 0 & 0 \\
		0 & 0 & 0 & 0 & 0 & e^{i\phi} \\
		0 & 0 & 1 & 0 & 0 & 0 \\
		0 & e^{i\phi} & 0 & 0 & 0 & 0 \\
		0 & 0 & 0 & 0 & 1 & 0 \\
		0 & 0 & 0 & e^{i\phi} & 0 & 0
	\end{pmatrix}
\end{align}
This matrix leaves one spatial mode (e.g., the lower arm) unaffected across time modes, but delays each time mode in the other spatial mode (the upper arm). For instance, it maps $a'_1 \rightarrow a'_2$, $a'_2 \rightarrow a'_3$. In a physical system, a pulse in the third time bin would be delayed to the next time bin. To maintain a finite-dimensional matrix representation, we model this as a cyclic shift, where the third mode is mapped back to the first ($a'_3 \rightarrow a'_1$). Since this third mode is virtual and assumed to be in the vacuum state, this mathematical convenience has no effect on the final results.

The operator for the second beam splitter, $BS_2$, is defined similarly to Eq.~\eqref{eq:BS} but with parameters $(T_2,R_2)$. The total unitary matrix $U_{MZ}$ for the MZI is then:

	\begin{align}
		U_{MZ} &= BS_2 \cdot D \cdot BS_1 \\
		&=
		\begin{pmatrix}
			\sqrt{T_2 T_1} & i \sqrt{T_2 R_1} & 0 & 0 & -\sqrt{R_2 R_1}e^{i\phi} & i\sqrt{R_2 T_1}e^{i\phi} \\
			i\sqrt{R_2 T_1} & -\sqrt{R_2 R_1} & 0 & 0 & i\sqrt{T_2 R_1}e^{i\phi} & \sqrt{T_2 T_1}e^{i\phi} \\
			-\sqrt{R_2 R_1}e^{i\phi} & i\sqrt{R_2 T_1}e^{i\phi} & \sqrt{T_2 T_1} & i \sqrt{T_2 R_1} & 0 & 0 \\
			i\sqrt{T_2 R_1}e^{i\phi} & \sqrt{T_2 T_1}e^{i\phi} & i\sqrt{R_2 T_1} & -\sqrt{R_2 R_1} & 0 & 0 \\
			0 & 0 & -\sqrt{R_2 R_1}e^{i\phi} & i\sqrt{R_2 T_1}e^{i\phi} & \sqrt{T_2 T_1} & i \sqrt{T_2 R_1} \\
			0 & 0 & i\sqrt{T_2 R_1}e^{i\phi} & \sqrt{T_2 T_1}e^{i\phi} & i\sqrt{R_2 T_1} & -\sqrt{R_2 R_1}
		\end{pmatrix}
	\end{align}

This matrix has three degrees of freedom: $T_1$, $T_2$, and $\phi$.

Our goal is to construct the bases $B_1 = \frac{1}{\sqrt{2}} (\sigma_z + \sigma_x)$ and $B_2 = \frac{1}{\sqrt{2}} (\sigma_z - \sigma_x)$. To realize a basis, its eigenvectors must produce deterministic outcomes. That is, for a given eigenvector input, one detector must fire while the other remains silent (due to constructive and destructive interference).

First, we construct the $B_1$ basis. Its eigenvectors are $|u_1 \rangle = \cos\frac{\pi}{8}|e \rangle + \sin\frac{\pi}{8}|l \rangle$ and $|u_2 \rangle = \sin\frac{\pi}{8}|e \rangle - \cos\frac{\pi}{8}|l \rangle$. Mapping these to our input vector representation $(a_1, a_2)^T$ and applying the transformation yields:
\begin{align}
	\begin{pmatrix}
		\cos\frac{\pi}{8} \\ 0 \\ \sin\frac{\pi}{8} \\ 0 \\ 0 \\ 0
	\end{pmatrix}
	\rightarrow
	\begin{pmatrix}
		\sqrt{T_2 T_1} \cos\frac{\pi}{8} \\
		i\sqrt{R_2 T_1} \cos\frac{\pi}{8} \\
		-\sqrt{R_2 R_1}e^{i\phi} \cos\frac{\pi}{8} + \sqrt{T_2 T_1} \sin\frac{\pi}{8} \\
		i\sqrt{T_2 R_1}e^{i\phi} \cos\frac{\pi}{8} + i\sqrt{R_2 T_1} \sin\frac{\pi}{8} \\
		-\sqrt{R_2 R_1}e^{i\phi} \sin\frac{\pi}{8} \\
		\sqrt{T_2 T_1}e^{i\phi} \sin\frac{\pi}{8}
	\end{pmatrix}
	\\
	\begin{pmatrix}
		\sin\frac{\pi}{8} \\ 0 \\ -\cos\frac{\pi}{8} \\ 0 \\ 0 \\ 0
	\end{pmatrix}
	\rightarrow
	\begin{pmatrix}
		\sqrt{T_2 T_1} \sin\frac{\pi}{8} \\
		i\sqrt{R_2 T_1} \sin\frac{\pi}{8} \\
		-\sqrt{R_2 R_1}e^{i\phi} \sin\frac{\pi}{8} - \sqrt{T_2 T_1} \cos\frac{\pi}{8} \\
		i\sqrt{T_2 R_1}e^{i\phi} \sin\frac{\pi}{8} - i\sqrt{R_2 T_1} \cos\frac{\pi}{8} \\
		+\sqrt{R_2 R_1}e^{i\phi} \cos\frac{\pi}{8} \\
		-\sqrt{T_2 T_1}e^{i\phi} \cos\frac{\pi}{8}
	\end{pmatrix}
\end{align}
For the input state $|u_1 \rangle$, we require detector 2 to click and detector 1 to be silent in the second time bin. For the input state $|u_2 \rangle$, the opposite must occur. This leads to the following system of equations:
\begin{align}
	-\sqrt{R_2 R_1} e^{i\phi} \cos\frac{\pi}{8} + \sqrt{T_2 T_1} \sin\frac{\pi}{8} = 0 \\
	i\sqrt{T_2 R_1} e^{i\phi} \sin\frac{\pi}{8} - i\sqrt{R_2 T_1} \cos\frac{\pi}{8} = 0
\end{align}
Solving these equations yields the conditions: $T_1 = R_1 = 0.5$, $\phi=0$, and $T_2 = \cos^2(\frac{\pi}{8})$. Thus, to construct the $B_1$ basis, the MZI parameters must be $(T_1,R_1)=(0.5,0.5)$, $\phi=0$, and $(T_2,R_2) \approx (0.853, 0.147)$.

Next, we construct the $B_2$ basis. Its eigenvectors are $|v_1 \rangle = \cos\frac{\pi}{8}|e \rangle - \sin\frac{\pi}{8}|l \rangle$ and $|v_2 \rangle = \sin\frac{\pi}{8}|e \rangle + \cos\frac{\pi}{8}|l \rangle$. Following the same procedure, we impose the condition that for input $|v_1 \rangle$, detector 2 clicks, and for $|v_2 \rangle$, detector 1 clicks. This gives the equations:
\begin{align}
	-\sqrt{R_2 R_1} e^{i\phi} \cos\frac{\pi}{8} - \sqrt{T_2 T_1} \sin\frac{\pi}{8} = 0 \\
	i\sqrt{T_2 R_1} e^{i\phi} \sin\frac{\pi}{8} + i\sqrt{R_2 T_1} \cos\frac{\pi}{8} = 0
\end{align}
Solving this system yields $T_1=R_1=0.5$, $\phi=\pi$, and $T_2 = \cos^2(\frac{\pi}{8})$. Therefore, to construct the $B_2$ basis, the MZI must be configured with $(T_1,R_1)=(0.5,0.5)$, $\phi=\pi$, and $(T_2,R_2) \approx (0.853, 0.147)$.


In summary, the modified MZI 
can implement both desired measurement bases. The splitting ratios of the beam splitters are kept constant, while the phase is randomly switched between $0$ and $\pi$. This allows Bob to randomly choose between measuring in the $B_1$ and $B_2$ basis.

\section{Calculation of the CHSH Correlation Terms}
\label{app:CHSH_calc}

To theoretically obtain the expectation values $\langle A_i B_j \rangle$, we need to calculate the click probabilities at each of Bob's detectors for the different states sent by Alice. The relevant click probabilities are determined during the second time bin where interference occurs. Based on the MZI configuration derived in Appendix~\ref{app:MZI_calc}, the output amplitudes at the two detectors in the interference time slot are given by:
\begin{align}
	b_1 &= \frac{1}{\sqrt{2}} \left( -e^{i\phi} \sin\frac{\pi}{8} \cdot a_1 + \cos\frac{\pi}{8} \cdot a_2 \right) \\
	b_2 &= \frac{i}{\sqrt{2}} \left( e^{i\phi} \cos\frac{\pi}{8} \cdot a_1 + \sin\frac{\pi}{8} \cdot a_2 \right)
\end{align}
where $a_1$ and $a_2$ are the input amplitudes in the early and late time bins, respectively, and $\phi$ is the phase set by Bob. The probability of a click is the squared magnitude of the corresponding amplitude (e.g., $P_1 = |b_1|^2$). 

We adopt the convention that a click at detector 2 corresponds to the positive eigenvalue (+1) and a click at detector 1 corresponds to the negative eigenvalue (-1) of Bob's measurement outcome. Let $\alpha'$ be the amplitude of the coherent state after passing through the channel and Bob's MZI. We now proceed to calculate each expectation value.

\subsection{Calculation of \texorpdfstring{$\langle A_1 B_1 \rangle$}{<A1B1>}}
Alice prepares a state in the $Z$ basis ($A_1$) and Bob measures in the $B_1$ basis (corresponding to $\phi=0$).

\paragraph{Case 1: Alice sends $|0_z\rangle$ (eigenvalue +1).}
The input amplitudes are $a_1 = \alpha'$ and $a_2 = 0$. The output amplitudes at Bob's side are:
\begin{align*}
	b_1 = -\frac{\alpha'}{\sqrt{2}} \sin\frac{\pi}{8}, \quad
	b_2 = i\frac{\alpha'}{\sqrt{2}} \cos\frac{\pi}{8}
\end{align*}
The corresponding probabilities for Bob's outcomes are:
\begin{align*}
	p_{11}^{+1,+1} &= |b_2|^2 = \frac{|\alpha'|^2}{2} \cos^2\frac{\pi}{8} \\
	p_{11}^{+1,-1} &= |b_1|^2 = \frac{|\alpha'|^2}{2} \sin^2\frac{\pi}{8}
\end{align*}

\paragraph{Case 2: Alice sends $|1_z\rangle$ (eigenvalue -1).}
The input amplitudes are $a_1 = 0$ and $a_2 = \alpha'$. The output amplitudes are:
\begin{align*}
	b_1 = \frac{\alpha'}{\sqrt{2}} \cos\frac{\pi}{8}, \quad
	b_2 = i\frac{\alpha'}{\sqrt{2}} \sin\frac{\pi}{8}
\end{align*}
And the probabilities are:
\begin{align*}
	p_{11}^{-1,+1} &= |b_2|^2 = \frac{|\alpha'|^2}{2} \sin^2\frac{\pi}{8} \\
	p_{11}^{-1,-1} &= |b_1|^2 = \frac{|\alpha'|^2}{2} \cos^2\frac{\pi}{8}
\end{align*}

The expectation value is calculated as:
\begin{align}
	\langle A_1 B_1 \rangle &= \frac{p_{11}^{+1,+1} - p_{11}^{+1,-1} - p_{11}^{-1,+1} + p_{11}^{-1,-1}}{p_{11}^{+1,+1} + p_{11}^{+1,-1} + p_{11}^{-1,+1} + p_{11}^{-1,-1}} \nonumber \\
	&= \frac{\frac{|\alpha'|^2}{2} \left[ \cos^2\frac{\pi}{8} - \sin^2\frac{\pi}{8} - \sin^2\frac{\pi}{8} + \cos^2\frac{\pi}{8} \right]}{\frac{|\alpha'|^2}{2} \left[ \cos^2\frac{\pi}{8} + \sin^2\frac{\pi}{8} + \sin^2\frac{\pi}{8} + \cos^2\frac{\pi}{8} \right]} \nonumber \\
	&= \frac{2 \left[ \cos^2\frac{\pi}{8} - \sin^2\frac{\pi}{8} \right]}{2 \left[ \cos^2\frac{\pi}{8} + \sin^2\frac{\pi}{8} \right]} = \cos\frac{\pi}{4} = \frac{1}{\sqrt{2}}
\end{align}

\subsection{Calculation of \texorpdfstring{$\langle A_1 B_2 \rangle$}{<A1B2>}}
Alice prepares in the $Z$ basis ($A_1$) and Bob measures in the $B_2$ basis ($\phi=\pi$, so $e^{i\phi}=-1$).

\paragraph{Case 1: Alice sends $|0_z\rangle$ (eigenvalue +1).}
With $a_1=\alpha'$ and $a_2=0$, the output amplitudes are:
\begin{align*}
	b_1 = \frac{\alpha'}{\sqrt{2}} \sin\frac{\pi}{8}, \quad b_2 = -i\frac{\alpha'}{\sqrt{2}} \cos\frac{\pi}{8}
\end{align*}
This leads to probabilities:
\begin{align*}
	p_{12}^{+1,+1} &= |b_2|^2 = \frac{|\alpha'|^2}{2} \cos^2\frac{\pi}{8} \\
	p_{12}^{+1,-1} &= |b_1|^2 = \frac{|\alpha'|^2}{2} \sin^2\frac{\pi}{8}
\end{align*}

\paragraph{Case 2: Alice sends $|1_z\rangle$ (eigenvalue -1).}
With $a_1=0$ and $a_2=\alpha'$, the output amplitudes are:
\begin{align*}
	b_1 = \frac{\alpha'}{\sqrt{2}} \cos\frac{\pi}{8}, \quad b_2 = i\frac{\alpha'}{\sqrt{2}} \sin\frac{\pi}{8}
\end{align*}
This leads to probabilities:
\begin{align*}
	p_{12}^{-1,+1} &= |b_2|^2 = \frac{|\alpha'|^2}{2} \sin^2\frac{\pi}{8} \\
	p_{12}^{-1,-1} &= |b_1|^2 = \frac{|\alpha'|^2}{2} \cos^2\frac{\pi}{8}
\end{align*}
The expectation value is identical to the previous case:
\begin{align}
	\langle A_1 B_2 \rangle &= \frac{2 \left[ \cos^2\frac{\pi}{8} - \sin^2\frac{\pi}{8} \right]}{2 \left[ \cos^2\frac{\pi}{8} + \sin^2\frac{\pi}{8} \right]} \nonumber \\
	&= \cos\frac{\pi}{4} = \frac{1}{\sqrt{2}}
\end{align}

\subsection{Calculation of \texorpdfstring{$\langle A_2 B_1 \rangle$}{<A2B1>}}
Alice prepares in the $X$ basis ($A_2$) and Bob measures in the $B_1$ basis ($\phi=0$).

\paragraph{Case 1: Alice sends $|0_x\rangle$ (eigenvalue +1).}
With $a_1 = a_2 = \alpha'/\sqrt{2}$, the amplitudes are:
\begin{align*}
	b_1 = \frac{\alpha'}{2} (\cos\frac{\pi}{8} - \sin\frac{\pi}{8}), \quad
	b_2 = \frac{i\alpha'}{2} (\cos\frac{\pi}{8} + \sin\frac{\pi}{8})
\end{align*}
Resulting in probabilities:
\begin{align*}
	p_{21}^{+1,+1} &= |b_2|^2 = \frac{|\alpha'|^2}{4}\left[\cos\frac{\pi}{8} + \sin\frac{\pi}{8}\right]^2 \\
	p_{21}^{+1,-1} &= |b_1|^2 = \frac{|\alpha'|^2}{4}\left[\cos\frac{\pi}{8} - \sin\frac{\pi}{8}\right]^2
\end{align*}

\paragraph{Case 2: Alice sends $|1_x\rangle$ (eigenvalue -1).}
With $a_1 = \alpha'/\sqrt{2}$ and $a_2 = -\alpha'/\sqrt{2}$, the amplitudes are:
\begin{align*}
	b_1 = -\frac{\alpha'}{2} (\cos\frac{\pi}{8} + \sin\frac{\pi}{8}), \quad
	b_2 = \frac{i\alpha'}{2} (\cos\frac{\pi}{8} - \sin\frac{\pi}{8})
\end{align*}
Resulting in probabilities:
\begin{align*}
	p_{21}^{-1,+1} &= |b_2|^2 = \frac{|\alpha'|^2}{4}\left[\cos\frac{\pi}{8} - \sin\frac{\pi}{8}\right]^2 \\
	p_{21}^{-1,-1} &= |b_1|^2 = \frac{|\alpha'|^2}{4}\left[\cos\frac{\pi}{8} + \sin\frac{\pi}{8}\right]^2
\end{align*}
The expectation value is:
\begin{align}
	\langle A_2 B_1 \rangle &= \frac{2 \left[ (\cos\frac{\pi}{8} + \sin\frac{\pi}{8})^2 - (\cos\frac{\pi}{8} - \sin\frac{\pi}{8})^2 \right]}{2 \left[ (\cos\frac{\pi}{8} + \sin\frac{\pi}{8})^2 + (\cos\frac{\pi}{8} - \sin\frac{\pi}{8})^2 \right]} \nonumber \\
	&= \frac{4 \sin\frac{\pi}{8} \cos\frac{\pi}{8}}{2(\cos^2\frac{\pi}{8} + \sin^2\frac{\pi}{8})} = \sin\frac{\pi}{4} = \frac{1}{\sqrt{2}}
\end{align}

\subsection{Calculation of \texorpdfstring{$\langle A_2 B_2 \rangle$}{<A2B2>}}
Alice prepares in the $X$ basis ($A_2$) and Bob measures in the $B_2$ basis ($\phi=\pi$).

\paragraph{Case 1: Alice sends $|0_x\rangle$ (eigenvalue +1).}
With $a_1 = a_2 = \alpha'/\sqrt{2}$, the amplitudes are:
\begin{align*}
	b_1 = \frac{\alpha'}{2}(\sin\frac{\pi}{8} + \cos\frac{\pi}{8}), \quad
	b_2 = \frac{i\alpha'}{2}(-\cos\frac{\pi}{8} + \sin\frac{\pi}{8})
\end{align*}
Resulting in probabilities:
\begin{align*}
	p_{22}^{+1,+1} &= |b_2|^2 = \frac{|\alpha'|^2}{4}\left[\sin\frac{\pi}{8} - \cos\frac{\pi}{8}\right]^2 \\
	p_{22}^{+1,-1} &= |b_1|^2 = \frac{|\alpha'|^2}{4}\left[\sin\frac{\pi}{8} + \cos\frac{\pi}{8}\right]^2
\end{align*}

\paragraph{Case 2: Alice sends $|1_x\rangle$ (eigenvalue -1).}
With $a_1 = \alpha'/\sqrt{2}$ and $a_2 = -\alpha'/\sqrt{2}$, the amplitudes are:
\begin{align*}
	b_1 = \frac{\alpha'}{2}(\cos\frac{\pi}{8} - \sin\frac{\pi}{8}), \quad
	b_2 = -\frac{i\alpha'}{2}(\cos\frac{\pi}{8} + \sin\frac{\pi}{8})
\end{align*}
Resulting in probabilities:
\begin{align*}
	p_{22}^{-1,+1} &= |b_2|^2 = \frac{|\alpha'|^2}{4}\left[\cos\frac{\pi}{8} + \sin\frac{\pi}{8}\right]^2 \\
	p_{22}^{-1,-1} &= |b_1|^2 = \frac{|\alpha'|^2}{4}\left[\cos\frac{\pi}{8} - \sin\frac{\pi}{8}\right]^2
\end{align*}
The expectation value is:
\begin{align}
	\langle A_2 B_2 \rangle &= \frac{2 \left[ (\sin\frac{\pi}{8} - \cos\frac{\pi}{8})^2 - (\cos\frac{\pi}{8} + \sin\frac{\pi}{8})^2 \right]}{2 \left[ (\sin\frac{\pi}{8} - \cos\frac{\pi}{8})^2 + (\cos\frac{\pi}{8} + \sin\frac{\pi}{8})^2 \right]} \nonumber \\
	&= \frac{-4 \sin\frac{\pi}{8} \cos\frac{\pi}{8}}{2(\sin^2\frac{\pi}{8} + \cos^2\frac{\pi}{8})} = -\frac{1}{\sqrt{2}}
\end{align}

As a result, the value of the parameter $S$ is obtained as:
\begin{align*}
	S &= \langle A_1 B_1 \rangle + \langle A_1 B_2 \rangle + \langle A_2 B_1 \rangle - \langle A_2 B_2 \rangle \\
	&= \frac{1}{\sqrt{2}} + \frac{1}{\sqrt{2}} + \frac{1}{\sqrt{2}} - (-\frac{1}{\sqrt{2}}) \\
	&= 2\sqrt{2}.
\end{align*}

\section*{Funding Declaration}

The authors gratefully acknowledge the financial support of the Research Centre for
Quantum Engineering and Photonics Technology, Sharif
University of Technology for this project
(Grant No. 140400101).

\section*{Acknowledgements}
We would like to give special thanks to Mahdi Shokhmkar and Behzad Azadi Faraz for their useful discussions.

\section*{Data Availability}
The Python codes and the simulated datasets generated during the current study are publicly available in the Zenodo repository at: [https://doi.org/10.5281/zenodo.20391865].

\bibliography{sample}

\end{document}